\documentclass[prb,amsmath,amssymb,superscriptaddress,twocolumn]{revtex4}
\usepackage{graphicx,color}
\usepackage{bbm}

\newcommand{\br}{{\bf r}}
\newcommand{\bx}{{\bf x}}
\newcommand{\by}{{\bf y}}
\newcommand{\bz}{{\bf z}}
\newcommand{\bk}{{\bf k}}

\newcommand{\bv}{{\bf v}}
\newcommand{\bq}{{\bf q}}
\newcommand{\bA}{{\bf A}}
\newcommand{\bB}{{\bf B}}
\newcommand{\bI}{{\bf I}}
\newcommand{\bG}{{\bf G}}

\newcommand{\bQ}{{\bf Q}}
\newcommand{\bR}{{\bf R}}
\newcommand{\bS}{{\bf S}}

\newcommand{\deltav}{\boldsymbol{\delta}}
\newcommand{\eps}{\epsilon}

\DeclareMathAlphabet{\mathpzc}{OT1}{pzc}{m}{it}
\DeclareMathOperator{\Di}{Di}

\begin{document}
\title{Relaxation of Nuclear Magnetic Moments and Site-Selective NMR in $d$-Wave Superconductors}
\author{Robert E. Throckmorton}
\author{Oskar Vafek}
\affiliation{National High Magnetic Field Laboratory and Department
of Physics,\\ Florida State University, Tallahassee, Florida 32306,
USA}
\date{\today}
\begin{abstract}
A new mechanism for relaxing the nuclear magnetic moments, in which a pair
of spin-polarized BCS quasiparticles, is emitted or absorbed, and which
dominates at low temperature, is identified in type-II $d$-wave
superconductors in an external magnetic field above $H_{c1}$. The
results of the theory are compared with the NMR experiments on YBCO
in high magnetic fields and found to agree without invoking
antiferromagnetic order in the vortex core.
\end{abstract}
\maketitle
\section{Introduction}
Being a bulk real-space probe, with information about the precession
and relaxation rates of nuclear spins at different sites inside the
sample, nuclear magnetic resonance (NMR) has served as one of the
key experimental tools\cite{Walstedt2008} in the study of the
electronic properties of high temperature cuprate superconductors
(HTS). By and large, the $^{17}$O NMR
data\cite{Curro2000,MitrovicNature,kakuyanagi2002,MitrovicPRB2003}
on HTS, in moderate to large magnetic fields, has been interpreted
as evidence for antiferromagnetic (AF) order in the vortex
core\cite{MitrovicPRB2003,Kivelson2003,PALee2006}.  This interpretation
was based on the dependence of the spin-lattice relaxation rate, $1/T_1$,
for the Cu nuclei in the normal state; the rate remains constant with
temperature\cite{Takigawa1991}, as opposed to being linear in $T$, as predicted by the
Korringa law\cite{Slichter1996}.

We reexamine this interpretation using analytical and numerical
solutions of the the Bogoliubov-de~Gennes equations and find that
the present data can be understood quantitatively without invoking
AF ordering. In particular, the low temperature upturn in $1/(T_1T)$
near the vortex core, but not away from it, can be understood to be
caused by a combination of two effects: 1) the increase of the
quasiparticle (qp) wavefunction near the core and 2) the shift of
the minimum of the qp band with spin along the applied magnetic
field (spin up) to negative energies due to Zeeman coupling. As a
result, a new electronic channel opens up for relaxing the nuclear
spin whereby a {\it pair} of spin up quasiparticles is emitted or absorbed. This is in
contrast to the standard "spin-flip" channel, in which a spin
up(down) qp is destroyed and a spin down(up) qp is created. At
temperatures below the Zeeman scale the latter channel freezes out
since the number of qp's with spin anti-aligned with the $B$-field
becomes thermally activated, and the former channel dominates. In
addition, we find that the broad NMR lineshape appears even in the
extreme type-II limit, where the diamagnetic response of the
$H$-field induced supercurrents can be neglected. The broadening here is
found to be due to the spatially non-uniform {\it paramagnetic} response of
$d$-wave superconductors in the vortex state.

Several other theoretical investigations of NMR in the mixed state
of the cuprate superconductors have been carried out.  One of the
earliest treatments was that of Takigawa {\it
et.al.}\cite{takigawa1999} using a self-consistent method of
solution of the Bogoliubov-de Gennes equations due to Wang and
MacDonald\cite{Wang1995}.  They found that $1/T_1$ is linear in
temperature near the vortex cores at low temperatures and
exhibits a small, Hebel-Slichter-like, peak near the superconducting
transition temperature.  At low temperature, the rates near the core
are also found to be  larger than the rates away from it, which approach
the usual $T^3$ dependence.  NMR in the $d$-wave mixed state was also
studied using a semiclassical approach\cite{WortisPRB2000}, and using
a linearized form of the Bogoliubov-de Gennes equations\cite{Knapp2002}.
The results of the linearized model give faster rates near the vortex cores
than away from them.  They found that $1/(T_1T)$ near the core increases
slowly with temperature up to $30\text{ K}$, and remains almost constant
over the same temperature range away from the core.  Importantly, these
works focused on the quasiparticle spin-flip channel, but, as mentioned
above and as we show in more detail below, the Zeeman coupling of the
quasiparticles, which cannot be ignored at large magnetic fields,
introduces an additional channel for spin-lattice relaxation which is
found to dominate at low temperature.

Our paper is organized as follows.  In Sec. II, we state the full
Hamiltonian of our system and review the basic formulas for the Knight
shift and $1/T_1T$.  In Sec. III, we introduce our model for the electronic
contribution to the Hamiltonian, derive the formulas for the Knight
shift and $1/T_1T$ for this model, and present the results of our
calculations.  In Sec. IV, we discuss the possible influence of
antiferromagnetic correlations.  Finally, we present
our conclusions in Sec. V.

\section{The Knight Shift and Spin-Lattice Relaxation Rate}

\subsection{Basic Model}
We will start by stating the Hamiltonian describing our system.
\begin{equation}
{\hat H}={\hat H}_e+{\hat H}_n+{\hat H}_{hf},
\end{equation}
where ${\hat H}_e$ is the Hamiltonian of the electrons on their own,
${\hat H}_n$ is that of the nuclei on their own, and ${\hat H}_{hf}$ is
the hyperfine interaction between the electrons and nuclei.  The nuclear
contribution is just the total energy of the nuclear spins in an applied
magnetic field,
\begin{equation}
{\hat H}_n=-\gamma_n\hbar\sum_{\br}{\hat\bI}(\br)\cdot\bB,
\end{equation}
where $\gamma_n$ is the gyromagnetic ratio of the nuclei and ${\hat\bI}(\br)$
is the spin of the nucleus at $\br$.  There are other terms present, such as
quadrupole terms and interactions among the nuclei\cite{Walstedt2008,Slichter1996,MitrovicThesis}.
However, in large magnetic fields, which we will be considering here, these terms
are small compared to the above magnetic term.  The quadrupole terms lead to uneven
splitting of the nuclear energies, and the interactions may lead to a slight broadening
of the resonances of the nuclei\cite{Slichter1996,MitrovicThesis}. We use the
model of Shastry, Mila, and Rice\cite{Shastry1989,MilaRice1989} for the hyperfine interaction,
\begin{equation}
{\hat H}_{hf}=-\gamma_e\gamma_n\hbar^2\sum_{\br\br'}{A(\br-\br')\hat{\bI}(\br)\cdot\hat{\bS}(\br')},
\end{equation}
where $\gamma_e$ and $\gamma_n$ are the gyromagnetic ratios of an electron
and a nucleus, respectively, $\hat{\bI}$ and $\hat{\bS}$ are their
respective spin angular momenta, and the coefficients $A(\br)$ are the form
factors for the hyperfine interaction\cite{Shastry1989,MilaRice1989}.  Note
that we are taking $\gamma_e$ to have a negative value.  The electronic
contribution will be the subject of the next section.

To compare our results to
experiments\cite{MitrovicPRB2003,MitrovicThesis}, we will be
interested in the relaxation rates and the Knight shifts for
the in-plane $^{17}$O atoms in YBCO.  For these atoms, we include the
contributions to the form factor $A(\br)$ from both
nearest- (n.n.) and next-nearest-neighbor (n.n.n.) copper atoms:
$\gamma_e\gamma_n\hbar^2 A(\br-\br')$ is equal to
$2.317\times 10^{-7}\text{ eV}$ for the n.n. Cu atoms, and
$5.794\times 10^{-8}\text{ eV}$ for the n.n.n. Cu
atoms\cite{MillisPRB1990,Zha1996}. As discussed by Zha, Barzykin,
and Pines\cite{Zha1996}, this form factor suppresses contributions from
the AF correlations to the spin-lattice relaxation rate at O-sites
in the normal state. We expect this suppression to persist in the
mixed state, as we will argue in Section IV.

\subsection{The Knight Shift}
The Knight shift is a change (usually an increase) in the nuclear
resonance frequency induced by the surrounding electrons\cite{Slichter1996}.
This can be attributed to an effective magnetic field produced by
the electrons through the hyperfine coupling to the nucleus.  Using
first-order time-independent perturbation theory on this term and
taking the thermal average of the result, we get
\begin{equation}
{\hat H}_{hf, eff}=-\gamma_n\hbar\sum_{\br}{\hat{\bI}(\br)\cdot\delta\bB_{eff}(\br)},
\end{equation}
where
\begin{equation}
\delta\bB_{eff}(\br)=\gamma_e\hbar\sum_{\br'}{A(\br-\br')\left <\hat{\bS}(\br')\right >}
\end{equation}
is the effective magnetic field experienced by the nuclei and produced
by the electrons; $\left <.\right >$ denotes a thermal average.
Because the nuclear resonance frequency, $\omega=\gamma_n B$, is proportional to the
applied magnetic field, this means that the resonance frequency is
shifted by an amount $\gamma_n\delta B_{eff}$.

We will use this formula in the next section to determine the Knight
shift in a $d$-wave superconductor in a magnetic field.  Note that the
above formulas, in general, allow for a position dependence of the
effective magnetic field; the Knight shift in our system will, in fact,
be position dependent.

\subsection{The Spin-Lattice Relaxation Rate}
As is well known\cite{Slichter1996}, the spin-lattice relaxation
rate at $\br$ is given by
\begin{equation}\label{1T1}
\frac{1}{T_1(\br)} = \tfrac{1}{2}\frac{\sum_{mn}W_{mn}(\br)(E_m -
E_n)^2}{\sum_{n}E_n^2},
\end{equation}
where $E_n$ is the energy of the nucleus at $\br$ in a state $n$,
and it is assumed that $\sum_{n}E_n = 0$.  We will also assume that
the energies are equally spaced, i.e. $E_n-E_{n-1} = \delta E$.
This is not exactly true, due to, for example, the quadrupole term,
but, because we are working in high magnetic fields, such contributions
beyond the magnetic energy are small, and we may treat the eigenstates
of this term alone as almost exact, which is one of the assumptions
made in the use of this formula\cite{Slichter1996}.

The function $W_{mn}(\br)$ entering Eq.(\ref{1T1}) is the transition
rate for the $z$ component (along the H-field) of the nuclear spin
at site $\br$ to go from $m\hbar$ to $n\hbar$. We can find these
rates using Fermi's Golden Rule,
\begin{equation}
W_{mn}(\br)=\frac{2\pi}{\hbar}\left <\sum_{QQ'}\left |\left
<mQ'\right |\hat{V}(\br)\left |nQ\right >\right |^2\delta(E_{mQ'}-
E_{nQ})\right >, \label{GoldenRule}
\end{equation}
where $V(\br)$ is the hyperfine interaction in the form,
\begin{equation}
V(\br)=-\gamma_e\gamma_n\hbar^2\sum_{\br'}{A(\br-\br')\hat{\bI}(\br)\cdot\hat{\bS}(\br')}.\label{HFIntV}
\end{equation}

We will employ these formulas to determine the spin-lattice relaxation
rate in a $d$-wave superconductor in a magnetic field in the next section.
Again, note that our formulas, in general, allow for a position dependence
of the relaxation rate and the Knight shift.

\section{NMR in a $d$-wave Superconductor}
\subsection{The Bogoliubov-de Gennes (BdG) Equation}
We now discuss the electronic contribution to the Hamiltonian.  Our
starting point will be the Hamiltonian for electrons on a square
tight-binding lattice in a magnetic field with a singlet pairing
term
\begin{eqnarray}
{\hat H}\!\!&=&\!\!\!\sum_{\left <\br\br'\right >}\left
[t_{\br\br'}{\hat c}_{\br\sigma}^{\dag}{\hat
c}_{\br'\sigma}+\Delta_{\br\br'}\left ({\hat
c}_{\br\uparrow}^{\dag}{\hat c}_{\br'\downarrow}^{\dag} - {\hat
c}_{\br\downarrow}^{\dag}{\hat
c}_{\br'\uparrow}^{\dag}\right )+\text{h.c.}\right ]\nonumber\\
&-&\sum_{\br}{{\hat
c}_{\br\alpha}^{\dag}\left(\mu\delta_{\alpha\beta}+h\sigma_{\alpha\beta}^z\right){\hat
c}_{\br\beta}}
\end{eqnarray}
where the tight-binding hopping constants, $t_{\br\br'}$ are
$t_{\br\br'}=-t e^{-i A_{\br\br'}}, \mathcal{A}_{\br\br'}=\frac{e}{\hbar
c}\int_{\br}^{\br'}\bA(\br)\cdot d\br$; $\bA(\br)$ is the vector
potential associated with the (constant) applied magnetic field
$\bB$, $\frac{1}{2}g\mu_B$ is the spin magnetic moment of an
electron, and $\mu$ is the chemical potential. For the symmetric
gauge $\bA(\br)=\frac{1}{2}\bB\times\br$, the values of
$\mathcal{A}_{\br\br'}$ relevant for a square lattice are
$\mathcal{A}_{\br,\br+{\hat\bx}}=-\pi y\Phi / \Phi_0$ and
$\mathcal{A}_{\br,\br+{\hat\by}}=\pi x\Phi / \Phi_0$, where $\Phi$ is the
magnetic flux through a plaquette, and $\Phi_0 = hc/e$ is the flux
quantum. The pairing field $\Delta_{\br\br'}$ is assumed to have a
constant magnitude and $\Delta_{\br\br'} = \eta_{\deltav}\Delta_0
e^{i\theta_{\br\br'}}$, where $\eta_{\pm {\hat\bx}}=-\eta_{\pm
{\hat\by}}=1$, otherwise $\eta_{\deltav}=0$. The phase factor $
e^{i\theta_{\br\br'}}=\frac{e^{i\phi(\br)} + e^{i\phi(\br')}}{\left
|e^{i\phi(\br)} + e^{i\phi(\br')}\right |}, $ where $\phi(\br)$
satisfies the equations, $\nabla\times\nabla\phi =
2\pi{\hat\bz}\sum_i{\delta(\br-\br_i)}$, where the $\br_i$ are the
positions of the vortex cores, and $\nabla^2\phi = 0$.  These
conditions determine $\phi(\br)$ up to terms of the form
$\phi_0+\bv_0\cdot\br$; these constants are fixed by requiring zero
overall current.  The vortex cores form a periodic Abrikosov
lattice, such that each primitive cell (magnetic unit cell) of this
lattice carries exactly one quantum of magnetic flux $hc/e$.  Note
that our assumption of a constant magnetic field effectively
corresponds to an infinite penetration depth.  We choose to assume
a pairing field of constant amplitude, placing all of the vortex physics
into the phase.  We do so because we believe that assuming a constant amplitude, as
opposed to calculating it self-consistently, will not greatly affect our
results.  Lacking a microscopic model for cuprate superconductors,
it is uncertain whether a self-consistent calculation will
result in much improvement of our results.  Finally, for notational convenience we
denote the Zeeman factor by $h=\frac{1}{2}g\mu_BB$.

Our method of solution for this problem follows Refs.\cite{vmftvmt,vm,MelikyanZBT2006}.
To diagonalize this Hamiltonian, we introduce the singular gauge-Bogoliubov de~Gennes
transformation\cite{ft,vm}
\begin{equation}
\begin{bmatrix}
{\hat c}_{\uparrow}(\br) \\
{\hat c}_{\downarrow}^{\dag}(\br)
\end{bmatrix}
= \sum_{\bk n}{
\begin{bmatrix}
e^{\frac{i}{2}\phi_{\br}}u_{\bk n}(\br) && -e^{\frac{i}{2}\phi_{\br}}v_{\bk n}^{\ast}(\br) \\
e^{-\frac{i}{2}\phi_{\br}}v_{\bk n}(\br) && e^{-\frac{i}{2}\phi_{\br}}u_{\bk n}^{\ast}(\br)
\end{bmatrix}
\begin{bmatrix}
{\hat\gamma}_{\bk n\uparrow} \\
{\hat\gamma}_{\bk n\downarrow}^{\dag}
\end{bmatrix}
}, \label{BogoliubovTrans}
\end{equation}
which allows us to rewrite the Hamiltonian in terms of the
quasiparticles in the Bloch basis corresponding to the magnetic unit
cell $\ell_x\times\ell_y$ containing a pair of vortices.  By Bloch's
theorem the Nambu spinors, which are eigenfunctions of the
Bogoliubov-de~Gennes equation\cite{ft,vmftvmt,vm}, can be written as
$[u_{\bk n}(\br),v_{\bk n}(\br)]^T=e^{i\bk\cdot\br}[U_{\bk
n}(\br),V_{\bk n}(\br)]^T$, where $U_{\bk n}(\br)$ and $V_{\bk
n}(\br)$ are periodic in $\ell_x\times\ell_y$, $n$ is the band index
and the crystal momentum $\bk\in
(-\frac{\pi}{\ell_x},\frac{\pi}{\ell_x}]\times(-\frac{\pi}{\ell_y},\frac{\pi}{\ell_y}]$.

\begin{figure}[tbh]
\centering
\includegraphics[width=\columnwidth,clip]{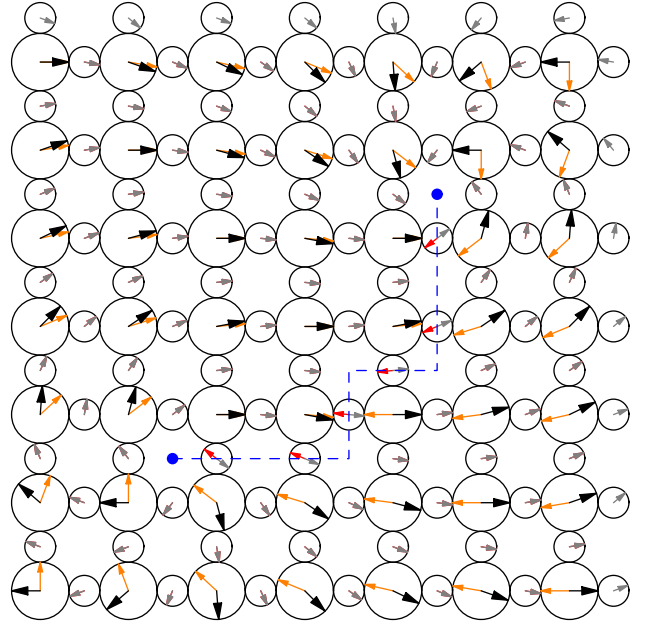}
\caption{\label{phasefactors}Illustration of the process by which we choose the values of $e^{i\phi(\br)/2}$ for each site for a $6\times 6$ magnetic unit cell.  The large circles represent the sites, and the small circles the bonds connecting them.  The blue dots are the locations of the vortex cores, and the dashed blue line is the branch cut.  The black arrows represent the site phase factors, $e^{i\phi(\br)}$, while the orange arrows represent the value of $e^{i\phi(\br)/2}$ chosen by the process outlined in the text.  The bond phase variable, $e^{i\theta_{\br\br'}}$, is represented by the gray arrow, and $e^{i\phi(\br)/2}e^{i\phi(\br')/2}$ is represented by the red arrow.
}
\end{figure}

There is one issue introduced by this transformation that is worth addressing
in detail. As we wind around a vortex, $\phi(\br)$ increases by $2\pi$.
This means that the phase factors in the above gauge transformation only
increase by $\pi$, meaning that the phase factors, at this point, are not
uniquely determined.  We must therefore introduce branch cuts into the
Hamiltonian and choose the values of the phase factors carefully. The
procedure we use in choosing the values of these factors is that used
in Ref.\cite{vm}.  We first choose a branch cut, which
can be any continuous curve connecting the two vortices inside the
magnetic unit cell.  We then choose one point $\br_0$ on the atomic
lattice, and let the phase factor for that site be $b_0 = e^{i\phi(\br_0)/2}$.
We now move to a neighboring site $\br$ such that we do not need to
cross the branch cut to reach it.  Let $b$ be the phase factor for this
site.  The solution to $b^2 = e^{i\phi(\br)}$ that we choose is the one
that gives the lower value of $\left |b - b_0\right |$.  We do this for
all sites, thus generating the appropriate values for $e^{i\phi(\br)/2}$.
This process is illustrated in Figure \ref{phasefactors}.

The coefficients $U_{\bk n}(\br)$ and $V_{\bk n}(\br)$ satisfy the
Bogoliubov-de Gennes equation, $e^{-i\bk\cdot\br}{\mathcal{\hat H}}_0e^{i\bk\cdot\br}\Phi_{\bk n}(\br)=E_{\bk n}\Phi_{\bk n}(\br)$,
where $\Phi_{\bk n}(\br)=[U_{\bk n}(\br),V_{\bk n}(\br)]^T$,
$\mathcal{\hat H}_0=\sigma_z(\mathcal{\hat E}_{\br}-\mu)+\sigma_x{\hat\Delta}_{\br}-h$,
the operators $\mathcal{\hat E}_{\br}$ and ${\hat\Delta}_{\br}$ are
\begin{eqnarray*}
\mathcal{\hat E}_{\br} &=& -t\sum_{\deltav = \pm{\hat\bx},\pm{\hat\by}}{z_{2,\br,\br + \deltav} e^{i\sigma_z V_{\br,\br + \deltav}}{\hat T}_{\deltav}} \\
{\hat\Delta}_{\br} &=& \Delta_0\sum_{\deltav = \pm{\hat\bx},\pm{\hat\by}}{z_{2,\br,\br + \deltav}\eta_{\deltav}{\hat T}_{\deltav}},
\end{eqnarray*}
and ${\hat T}_{\deltav}$ performs a translation along the vector $\deltav$.

We now wish to make a comment on the energies of the quasiparticles in our system.  Let
$\mathcal{\hat H}_{0,NZ}=\sigma_z(\mathcal{\hat E}_{\br}-\mu)+\sigma_x{\hat\Delta}_{\br}$
--- that is, $\mathcal{\hat H}_{0,NZ}$ is $\mathcal{\hat H}_0$ without the Zeeman term.
Note that diagonalizing $\mathcal{\hat H}_{0,NZ}$ is the same as diagonalizing $\mathcal{\hat H}_0$
because the two differ only by a term proportional to the identity matrix.  In fact, if
we let $E_{\bk n}$ be the eigenvalues of $\mathcal{\hat H}_{0,NZ}$, then the eigenvalues
of $\mathcal{\hat H}_0$ are just $E=E_{\bk n}-h$.  The matrix $\mathcal{\hat H}_{0,NZ}$,
as we implied, would replace $\mathcal{\hat H}_0$ if we neglected the Zeeman splitting.
Since we can simultaneously diagonalize $\mathcal{\hat H}_0$ and $\mathcal{\hat H}_{0,NZ}$,
we see that the same eigenvectors would diagonalize the difference between the two Hamiltonians
that result in these matrices, which is a term proportional to the $z$ component of the
spin.  This is because the $z$ component of the spin is a good quantum number, and can be used
to label the elementary excitations.  We note that
\begin{eqnarray}
[(i\sigma_y)\mathcal{\hat H}_{0,NZ}(-i\sigma_y)]^{\ast}=\sigma_y\mathcal{\hat H}_{0,NZ}^{\ast}\sigma_y \cr
=-\sigma_z(\mathcal{\hat E}_{\br}-\mu)-\sigma_x{\hat\Delta}_{\br}=-\mathcal{\hat H}_{0,NZ}.
\end{eqnarray}
This shows that, if we multiply an eigenvector of $\mathcal{\hat H}_{0,NZ}$ by $i\sigma_y$
and then take the complex conjugate of the result, then we obtain another eigenvector of
the same matrix, but with the negative of the eigenvalue of the original vector.  This is
exactly what we did to obtain $\psi_{\bk n}(\br)$ from $\psi'(\br)$.  We have thus shown
that the spinor $\psi'(\br)$ is an eigenvector of $\mathcal{\hat H}_{0,NZ}$ with eigenvalue
$-E_{\bk n}$, and therefore also an eigenvector of $\mathcal{\hat H}_0$ with eigenvalue
$-E_{\bk n}-h$.

The fact that we can generate the negative-energy states from the positive-energy ones
implies that we may take the sum in equation \eqref{BogoliubovTrans} to be over states
that give positive eigenvalues of $\mathcal{\hat H}_{0,NZ}$.  We could also choose, for
example, the states with negative eigenvalues.  In fact, we may choose any one of these
two possibilities for each term.  In this paper, we conform to widely-used convention and
use the positive energy eigenvalues.

The diagonalized Hamiltonian takes on the form
\begin{eqnarray}
{\hat H} &=& \sum_{\bk n}(E_{\bk n\uparrow}{\hat\gamma}_{\bk
n\uparrow}^{\dag}{\hat\gamma}_{\bk n\uparrow} + E_{\bk
n\downarrow}{\hat\gamma}_{\bk n\downarrow}^{\dag}{\hat\gamma}_{\bk
n\downarrow})-E^{(0)},
\end{eqnarray}
where $E^{(0)}=N\mu+\sum_{\bk n}{E_{\bk n}}$ and the qp
eigenenergies are $E_{\bk n\sigma} = E_{\bk n} - \sigma h$. The
density of states (without Zeeman)
$N(\omega)=\sum^{\ell_x\ell_y}_{n=1}\int\frac{d^2\bk}{\Omega_{BZ}}\delta(\omega-E_{\bk
n})$ for realistic values of the physical parameters is plotted in
Fig. \ref{fig:DOS}.  As shown below, these energies and wavefunctions
enter into the calculation of the NMR line shape and $1/T_1$.
\begin{figure}[t]
\centering
\includegraphics[width=\columnwidth,clip]{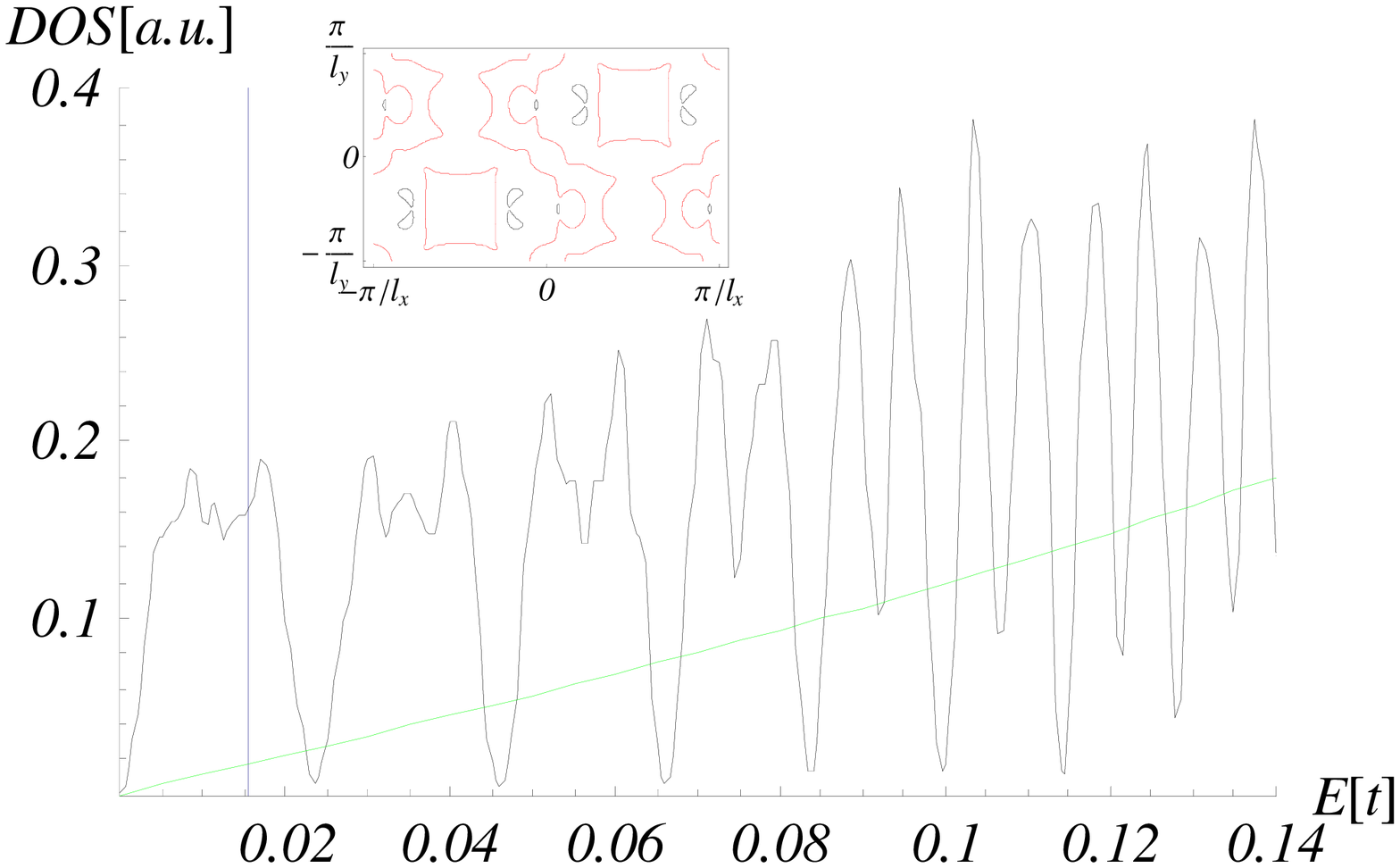}
\caption{\label{fig:DOS} Density of qp states for $B=0$ (green line)
and $B\approx 42$T ($\ell_x=20a$, $\ell_y=34a$; black line) for
$\alpha_D=14$ and $\mu=0.3t$. The vertical line shows the
corresponding Zeeman shift $h=\frac{1}{2}g\mu_BB$. The inset shows
the spin polarized qp Fermi surfaces which come from the lowest
(red) and the next to lowest (black) magnetic bands.}
\end{figure}

In our calculations, we worked with both $20\times 34$ and $26\times 26$ unit cells,
both of which correspond roughly to an applied field of $42\text{ T}$, as well as with
a $36\times 62$ unit cell, which corresponds to an applied field of about $13\text{ T}$.
These calculations were done for optimally-doped YBCO, for which $t = 153\text{ meV}$,
$\Delta_0 = \frac{1}{14}t$, and $\mu = 0.297t$.  Because we wish to calculate thermodynamic
properties of the system at low temperatures (temperatures up to $30\text{ K}$), we
only needed to find some of the lower energy bands.  Thus, we used the Arnoldi method
to find the energies and wave functions, and we discretized the reciprocal lattice
into a $50\times 50$ grid.

Let us now make some comments on the energy spectrum and the wave functions.  First,
we note that, while the energies $E_{\bk n}$ are all positive, so that no quasiparticles
would be present in the ground state of our system if there is no magnetic field, it
is possible, under certain circumstances, for some of the energies $E_{\bk n\uparrow}$
to be negative.  To be exact, $E_{\bk n}$ for most values of the chemical potential
could exhibit a gap.  If $E_{\bk n\uparrow}=E_{\bk n}-h$ is smaller than the Zeeman
splitting, then some of the energies will become negative\cite{vm}.  This means that,
in the ground state, there will be some quasiparticles present, all with their magnetic
moments parallel to the field.  This means that the ``gas'' of quasiparticles is spin-polarized in
the ground state, resulting in a non-zero Knight shift, even at zero temperature.

\subsection{The Knight Shift and Line Shape Broadening}
To find the Knight shift for a superconductor in a magnetic field, we simply substitute
the Bogoliubov transformation \eqref{BogoliubovTrans} into the spin operator.  Here, we
only consider the effect of the spin component along the $z$ axis,
\begin{equation*}
{\hat S}_z(\br) = {\hat c}_{\uparrow}^{\dag}(\br){\hat c}_{\uparrow}(\br) - {\hat c}_{\downarrow}^{\dag}(\br){\hat c}_{\downarrow}(\br).
\end{equation*}
Upon performing the Bogoliubov transformation and taking the thermal average, we find that
the effective magnetic field shift is
\begin{equation}
\delta\bB_{eff}(\br)=\gamma_e\hbar\sum_{\br'}\sum_{\bk n}{A(\br - \br')n_{\bk n}(\br')[f(E_{\bk n\downarrow}) - f(E_{\bk n\uparrow})]},
\end{equation}
where $f(E)$ is the usual Fermi-Dirac distribution,
\begin{equation}
f(E) = \frac{1}{e^{E/k_B T} + 1},
\end{equation}
and $n_{\bk n}(\br) = \left |u_{\bk n}(\br)\right |^2 + \left |v_{\bk n}(\br)\right |^2$.

\begin{figure}[t]
\centering
\includegraphics[width=\columnwidth,clip]{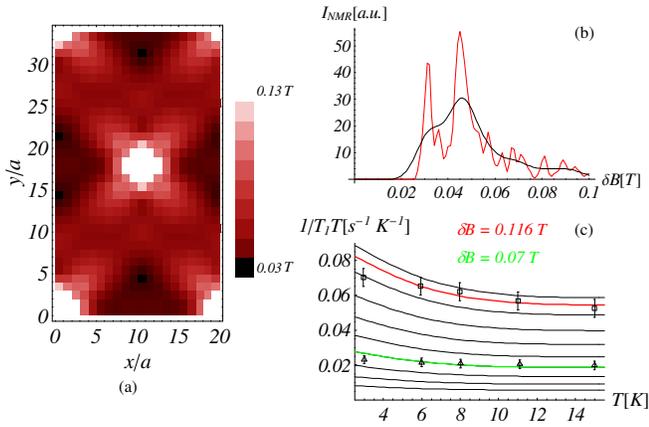}
\caption{\label{fig:MagLSRR} (a) Effective magnetic field shift
$\delta B_{eff}$ at $T=0$ as seen by the $^{17}$O nuclear spins for
(approximately) triangular vortex lattice corresponding to the
external field $B=42\text{ T}$. The Dirac anisotropy
$\alpha_D=t/\Delta_0=14$ and $\mu=0.3t$ corresponding to
$x\approx15\%$. (b) Spatial variation of $\delta B_{eff}$, whose
density is shown in red, leads to broadening of the NMR line shape
(black) (additionally broadened by a Gaussian with $\sigma=50$ gauss
\cite{MitrovicThesis}). (c) Spin-lattice relaxation rate $1/(T_1T)$
vs. $T$ for different $\delta B_{eff}$. The data points and error
bars are the experimental data\cite{MitrovicPRB2003}.}
\end{figure}

\begin{figure}[t]
\centering
\includegraphics[width=\columnwidth,clip]{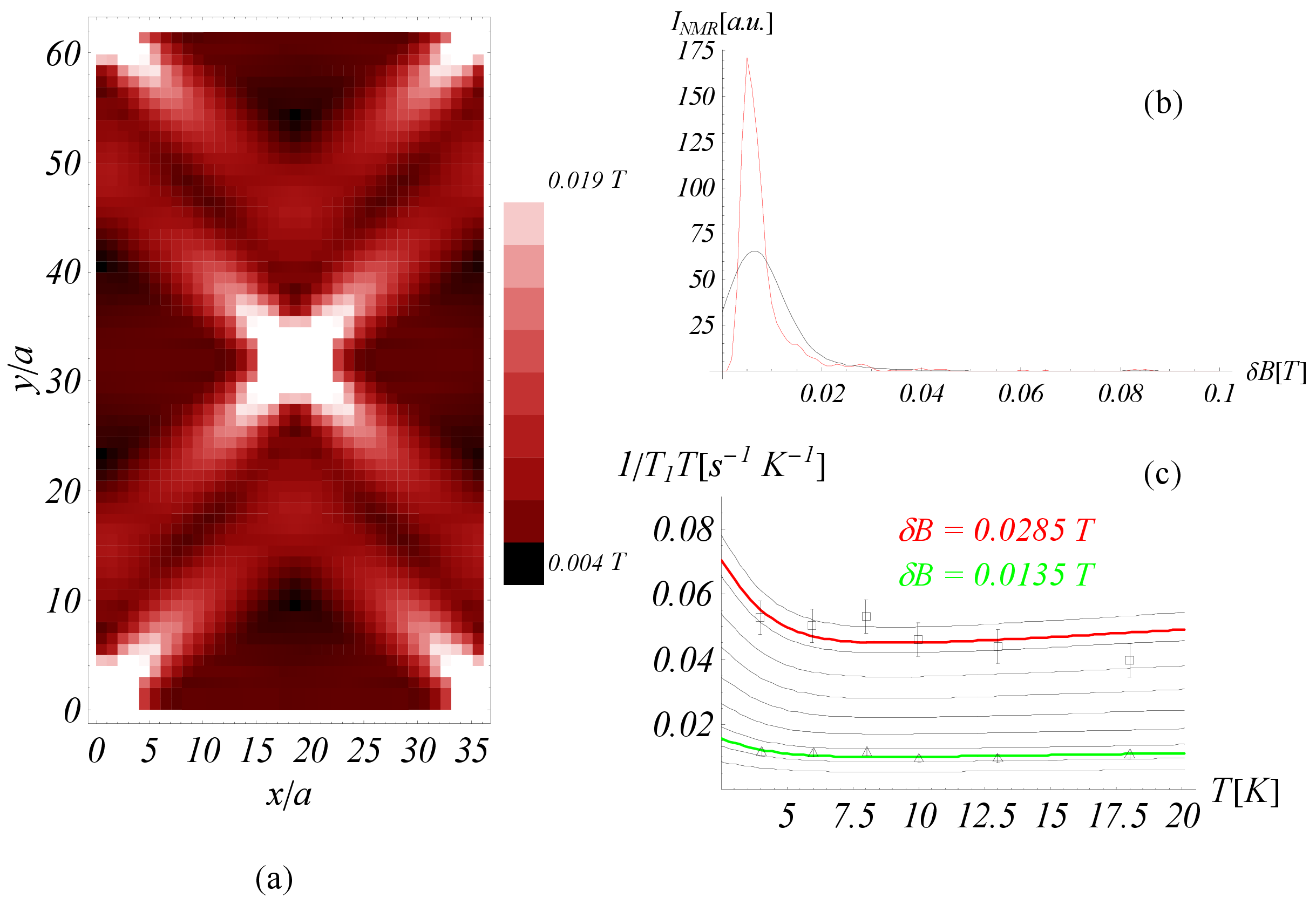}
\caption{\label{fig:MagLSRR2} (a) Effective magnetic field shift
$\delta B_{eff}$ at $T=0$ as seen by the $^{17}$O nuclear spins for
(approximately) triangular vortex lattice corresponding to the
external field $B=13\text{ T}$. All other parameters are the same
as before. (b) Spatial variation of $\delta B_{eff}$, whose
density is shown in red, leads to broadening of the NMR line shape
(black) (additionally broadened by a Gaussian with $\sigma=50$ gauss
\cite{MitrovicThesis}). (c) Spin-lattice relaxation rate $1/(T_1T)$
vs. $T$ for different $\delta B_{eff}$. Again, the data points and
error bars are the experimental data\cite{MitrovicPRB2003}.}
\end{figure}

Note that, even in the extreme type-II limit (i.e. taking
the penetration depth to infinity), in the vortex state the local
electron density is different at different locations within a
magnetic unit cell. Taking into account the Zeeman shift, this
translates to spatially varying spin density and by the above two
Eqs. to the spatially varying Knight shift. This means that even
nuclei of the same species will have different resonance frequencies
depending on their location in the magnetic unit cell. This results
in a broadening of the NMR line shape (Figures \ref{fig:MagLSRR} and
\ref{fig:MagLSRR2}).

We found the effective magnetic field due to the electrons, which
is proportional to the Knight shift, for the $20\times 34$ and
$36\times 62$ lattices, and we have plotted the spatial profiles
for this case in Figures \ref{fig:MagLSRR} and \ref{fig:MagLSRR2}.
As expected, the largest Knight shifts occur near the vortices
because the local electron density is largest in the same area.
In reality, the plots shown are for those O atoms on bonds parallel
to the ``short'' axis of the magnetic unit cell (in this case, the
$x$ axis); the plots for the atoms on bonds along the $y$ axis are
similar in appearance.

We now determine the line shape that would result from this effective
magnetic field.  Denoting this line shape by $f(B_0)$, the line shape
is given by
\begin{equation}
f(B_0)=\int{\delta[B_0 - B(\br)]\,d^3\br}.
\end{equation}
This formula assumes that the response of a single nucleus as a function
of the frequency is a delta function centered at the resonance frequency.
This, however, is not true in reality; in fact, the response has a finite
width in the frequency.  For this reason, we must convolute this line shape
with a broadening function representing the response of a single nucleus
to obtain the true line shape.  The broadening function we use in our
calculations is a Gaussian of width $50\text{ G}$.  This broadening is
experimentally motivated; the width is approximately that of the
$-1/2\leftrightarrow -3/2$ transition found by Mitrovi\'c\cite{MitrovicThesis}.
Again, we calculate this line shape for $20\times 34$ and $36\times 62$
lattices; the results are plotted in Figures \ref{fig:MagLSRR} and
\ref{fig:MagLSRR2}.

\subsection{Nuclear Spin-Lattice Relaxation Rate}
To find the nuclear spin-lattice relaxation rate, $1/T_1$, we first make a
simplification to the hyperfine interaction \eqref{HFIntV}.  Using the identity,
$\hat{I}_+\hat{S}_-+\hat{I}_-\hat{S}_+=2(\hat{I}_x\hat{S}_x+\hat{I}_y\hat{S}_y)$,
we may write it as
\begin{equation}
\hat{V}(\br)=-\gamma_e\gamma_n\hbar^2\sum_{\bR}{C(\br-\bR)[\hat{I}_{+}(\br)\hat{S}_{-}(\bR)+\hat{I}_{-}(\br)\hat{S}_{+}(\bR)]},
\end{equation}
where, for convenience, we have defined $C=\frac{1}{2}A$.  Note that we dropped
the $\hat{I}_z\hat{S}_z$ term; this term will only contribute when $m=n$, and
these transition rates, as we will see, do not contribute to the relaxation rate.
This form will be more convenient to work with because the $\hat{S}_{\pm}$ operators
take on simple forms, namely $\hat{S}_+=\hat{c}_{\uparrow}^{\dag}\hat{c}_{\downarrow}$
and $\hat{S}_-=\hat{c}_{\downarrow}^{\dag}\hat{c}_{\uparrow}$.

We now make an approximation.  We assume that the nuclear Zeeman energy is much
smaller than the electronic Zeeman energy, and thus we neglect that contribution
to the total energy of the system.  This is a good approximation if $E_m-E_n\ll k_BT$.
We then obtain
\begin{equation}
W_{mn}(\br)=\frac{2\pi}{\hbar}\left <\sum_{QQ'}\left |\left <mQ'\right |\hat{V}(\br)\left |nQ\right >\right |^2 \delta(E_{Q'}-E_{Q})\right >.
\end{equation}
Upon expanding out the expression, $\left |\left <mQ'\right |\hat{V}(\br)\left |nQ\right >\right |^2$,
occurring in equation \eqref{GoldenRule}, we obtain four terms; only
two of these will be non-zero, namely the term involving
$\left <m\right |\hat{I}_+\left |n\right >\left <n\right |\hat{I}_-\left |m\right >$
and the term of the same form, but with $m$ and $n$ interchanged.
Substituting this into equation \eqref{GoldenRule}, we get
\begin{eqnarray}
W_{mn}(\br)=\frac{2\pi}{\hbar}\gamma_e^2\gamma_n^2\hbar^4\left <\sum_{QQ'}\sum_{\bR\bR'}C(\br-\bR)C(\br-\bR')\right. \cr
\left (\left <m\right |\hat{I}_{+}(\br)\left |n\right >\left <n\right |\hat{I}_{-}(\br)\left |m\right >\left <Q'\right |\hat{S}_{-}(\bR)\left |Q\right >\left <Q\right |\hat{S}_{+}(\bR')\left |Q'\right >\right. \cr
\left.\left.+(m\leftrightarrow n, Q'\leftrightarrow Q)\right )\delta(E_{Q'}-E_{Q})\right >.
\end{eqnarray}
At this point, we express the spin raising and lowering operators in
terms of the quasiparticle operators using the above definitions and
equation \eqref{BogoliubovTrans}, and then introduce these operators
into the above expression.  Upon doing so, we obtain 16 terms.  However,
10 of these will be zero because they will involve expressions such as
$\left <Q'\right |\hat{\gamma}\hat{\gamma}\left |Q\right >\left <Q\right |\hat{\gamma}\hat{\gamma}^{\dag}\left |Q\right >$,
and it is impossible to ``match'' the operators in the first factor to
those in the second --- that is, we cannot pair, for example, an annihilation
operator in the first factor with its corresponding creation operator in
the second.  Of the six terms that remain, one of them, which has the form,
$\left <Q'\right |\hat{\gamma}_{\downarrow}\hat{\gamma}_{\downarrow}\left |Q\right >\left <Q\right |\hat{\gamma}_{\downarrow}^{\dag}\hat{\gamma}_{\downarrow}^{\dag}\left |Q'\right >$,
will also be zero because the process of creating or destroying two spin down
quasiparticles violates conservation of energy due to all spin down
quasiparticles having positive energy.  We note, however, that the
corresponding process for spin \textit{up} quasiparticles,  $\left <Q'\right |\hat{\gamma}_{\uparrow}\hat{\gamma}_{\uparrow}\left |Q\right >\left <Q\right |\hat{\gamma}_{\uparrow}^{\dag}\hat{\gamma}_{\uparrow}^{\dag}\left |Q'\right >$,
does \textit{not} violate conservation of energy because some of the
spin up quasiparticles have negative energies.  This means that, in
addition to the usual spin-flip scattering process (SF), there is also
a quasiparticle creation/annihilation process (PCA) through which the
nuclear spins can relax.

We go into detail on how we find the different terms occurring in our final
result in the Appendix; we only quote the final result here.  Performing the
appropriate sums and thermal averages we eventually obtain
$$
W_{mn}(\br) = 2\pi\gamma^2_e\gamma^2_n\hbar^3\left
[I^{mn}_{+}(\br)I^{nm}_{-}(\br)+ \text{c.c.}\right ] f(\br, T),
$$
where $I^{mn}_{+}(\br)=\left <m\right |I_+(\br)\left |n\right >$, and similarly
for $I^{mn}_{-}(\br)$.  This in turn gives the main result of this paper,
\begin{equation}\label{1T1_result}
\frac{1}{T_1(\br)} = 4\pi\gamma^2_e\gamma^2_n\hbar^3 f(\br, T),
\end{equation}
where the function $f(\br,T)=$
\begin{eqnarray}\label{eq:f}
&&
\sum_{nn'}\int\frac{d^2\bk}{\Omega_{BZ}}\frac{d^2\bk'}{\Omega_{BZ}}\left
[\frac{\left |G_{\bk n\bk' n'}^{\text{SF}}(\br)\right
|^2\delta(E_{\bk n} - E_{\bk' n'} + 2h)}{4\cosh^2\left(\frac{E_{\bk
n}+h}{2k_BT}\right)} \right. \nonumber\\
 &+& \left.\frac{\left |G_{\bk n\bk' n'}^{\text{PCA}}(\br)\right |^2\delta(E_{\bk n} + E_{\bk' n'} - 2h)}
 {8\cosh^2\left(\frac{E_{\bk n}-h}{2k_BT}\right)}
 \right].
\end{eqnarray}
The above integrals are over the $1^\text{st}$ Brillouin zone whose
area is $\Omega_{BZ}=4\pi^2/(\ell_x\ell_y)$. The qp coherence
factors enter via the functions $G_{\bk n\bk' n'}^{\text{SF}}(\br)=$
$$\sum_{\bR}C_{\br - \bR}\left [U_{\bk n}^{\ast}(\bR)U_{\bk' n'}(\bR)
+ V_{\bk n}^{\ast}(\bR)V_{\bk' n'}(\bR)\right ]e^{i(\bk' -
\bk)\cdot\bR} $$ and $G_{\bk n\bk' n'}^{\text{PCA}}(\br)=$
$$
\sum_{\bR}C_{\br - \bR}\left [V_{\bk n}(\bR)U_{\bk' n'}(\bR) -
U_{\bk n}(\bR)V_{\bk' n'}(\bR)\right ]e^{i(\bk' + \bk)\cdot\bR}.
$$
Note that this differs from the formulas presented in Refs.\cite{takigawa1999,Knapp2002},
most importantly by the presence
of the second term. The eigenenergies $E_{n\bk}\geq0$ are the
solutions of the Bogoliubov-de~Gennes equation without the Zeeman
coupling, and the corresponding periodic wavefunctions are
normalized within the magnetic unit cell: $\sum_{\br\in
\ell_x\ell_y}\left(|U_{\bk n}|^2(\br)+|V_{\bk n}|^2(\br)\right)=1$.
From Eqs. \eqref{1T1_result}-\eqref{eq:f} it is readily seen that,
regardless of the minimal value of $E_{n\bk}$, at temperatures $T\ll
h=\frac{1}{2}g\mu_BB$, the qp spin-flip (SF) process is activated
and thus vanishingly small. At $B=42\text{ T}$ this corresponds to
a temperature scale of $\sim28\text{ K}$, which in turn means that
the low $T$($\sim5\text{ K}$) upturn in $1/(T_1T)$ observed
experimentally\cite{MitrovicNature,MitrovicPRB2003} cannot be due
to this process. It is the second term (PCA) which dominates at
low temperatures and corresponds to the observed effect.

To illustrate the basic physics behind the effect, we will
temporarily ignore the orbital effects and analyze the consequences
of the Zeeman coupling alone\cite{yangsondhi1998}. Physically, this
would correspond to a thin film in a parallel (in-plane) B-field.
The eigenenergies in Eq.\eqref{eq:f} are then easily found to be
$E_{\bk}=\sqrt{\eps^2_{\bk}+\Delta^2_{\bk}}$, where
$\eps_{\bk}=-2t(\cos k_xa+\cos k_ya)-\mu$ and
$\Delta_{\bk}=2\Delta_0(\cos k_xa-\cos k_ya)$. At the same time the
wavefunctions are simply
$u_{\bk}=\frac{1}{\sqrt{2}}\sqrt{1+\frac{\eps_{\bk}}{E_{\bk}}}$ and
$v_{\bk}=\frac{\mbox{sgn}\Delta_{\bk}}{\sqrt{2}}\sqrt{1-\frac{\eps_{\bk}}{E_{\bk}}}$.
Assuming for simplicity $C_{\br}=c_0\delta_{\br,0}$ and
particle-hole symmetry, this gives for the Zeeman-only case
\begin{eqnarray}
\frac{1}{T_1}&=&\!2\pi\gamma^2_e\gamma^2_n\hbar^3c_0^2\times\left(\int_0^{\infty}dE\frac{N(E)N(E+2h)}{4\cosh^2\left(\frac{E+h}{2T}\right)}\right.\nonumber\\
&+&\left.
\int_0^{2h}dE\frac{N(E)N(2h-E)}{8\cosh^2\left(\frac{E-h}{2T}\right)}\right)
\end{eqnarray}
For $h,T\ll \Delta_0$ we need only the low energy qp density of
states, which is $N(E)=2E/(\pi v_Fv_{\Delta})$, where
$v_F=2\sqrt{2}at\sqrt{1-\frac{\mu^2}{16t^2}}$ and
$v_{\Delta}=v_F\Delta_0/t$. In this limiting case, the integral can
be found in a closed form and
\begin{eqnarray}
\frac{1}{T_1}&=&\frac{4}{\pi}\frac{\gamma^2_e\gamma^2_n\hbar^3c_0^2}{v^2_Fv^2_{\Delta}}\;
T^3F\left(\frac{h}{T}\right)
\end{eqnarray}
where $F(x)=\pi^2+8x\ln(1+e^x)-3x^2+8Li_2(-e^x)$ and $Li_s(z)$ is
the polylogarithm. For $x\ll1$, $F(x)=\frac{\pi^2}{3}-x^2$, and in
this limit we recover the standard d-wave $1/(T_1T)\sim T^2$. On the
other hand, for $x\gg1$, $F(x)=x^2-\frac{\pi^2}{3}$. In this limit
$1/(T_1T)$ increases as $T$ is lowered and approaches a constant at
$T=0$. The minimum in $1/(T_1T)$ then results from the competition
between the spin-flip process which dominates at $T\gg h$ and the qp
pair creation/annihilation process which dominates at $T\ll h$.

Putting back the coupling of the $B$-field to the orbital motion of
the electrons, we find that the effect described above acquires an
interesting spatial content. The dispersing states which are pulled
below zero energy by the Zeeman coupling are strongly concentrated
around the cores. Due to the increase of the low-energy wavefunctions
near the cores, the low-temperature relaxation rate of the nuclear
spin is the largest in the vicinity of the cores and {\it decreases}
with increasing $T$. This temperature dependence is in turn due to
the pair creation/annihilation processes, i.e the second term in
Equation \eqref{eq:f}.

To generate a dependence of the spin-lattice relaxation rate on the
internal magnetic field, we associated the rate at a given point and
a given temperature with the value of the effective magnetic field shift
at that point and temperature.  We then fit this set of points to a
power law to generate a continuous dependence; at all magnetic field shifts
of interest, the points are close enough together that they approximately
form a continuum.  We do this for all temperatures up to $30\text{ K}$ for
$B=42\text{ T}$ and up to $20\text{ K}$ for $B=13\text{ T}$; the results of
this procedure are shown in Figures \ref{fig:MagLSRR} and \ref{fig:MagLSRR2}.
We also highlight the curves that give the best fit to the data near the
vortex core and away from the core\cite{MitrovicNature,MitrovicPRB2003}.

\section{Contribution of Antiferromagnetc Correlations to the Spin-Lattice Relaxation Rate}
We now address the issue of how much of an effect antiferromagnetic
correlations will have on the spin-lattice relaxation rate of $^{17}$O
in the vortex state, assuming that the vortex cores represent normal-state
regions.  As was mentioned before, it is known that the form factor
filters out such correlations in the normal state\cite{Zha1996}.  Within the
simple model presented here, we find that this filtering is still active in the vortex state.  To
investigate the effect of vortices, we used a modification of the
phenomenological model set forth, among others, by Zha, Barzykin, and
Pines \cite{Zha1996}.  We start with their expression for the
``antiferromagnetic'' part of the susceptibility,
\begin{equation}
\chi_{\text{AF}}(\bk,\omega)=\frac{1}{4}\sum_i\frac{\alpha\xi^2\mu_B}{1+(\bq-\bQ_i)^2+i\omega/\omega_{\text{SC}}}.\label{AFsusc_simp}
\end{equation}
Here, $\alpha$ is a scale factor, $\xi$ is the antiferromagnetic correlation
length, $\mu_B$ is the Bohr magneton, the $\bQ_i$ are the locations
of the peaks in the susceptibility found from neutron scattering
experiments, $\omega_{\text{SC}}$ is the characteristic frequency of
spin fluctuations, and $\bq$ ranges over the entire first Brillouin
zone\cite{Zha1996}.  We obtained the model we used by rewriting the
above susceptibility as a function of position, separating the position
dependence into a dependence on the position of a ``magnetic unit cell''
and a dependence on position within the cell, and Fourier transforming
the result with respect to the magnetic unit cell position.  The result is
$\chi_{\text{AF}}(\bq,\delta\br-\delta\br',\omega)=$
\begin{equation}
\frac{1}{4}\frac{N_M}{N}\sum_{\bG}\sum_i\frac{\alpha\xi^2\mu_B e^{i(\bq+\bG)\cdot(\delta\br-\delta\br')}}{1+(\bq+\bG-\bQ_i)^2+i\omega/\omega_{\text{SC}}},
\end{equation}
where $\bq$ now ranges over the first magnetic Brillouin zone, $\bG$ is
the set of all vectors such that $e^{i\bG\cdot\bR}=1$ for all $\bR$ in the
magnetic lattice and such that $\bq+\bG$ is within the first atomic Brillouin
zone, $N_M$ is the number of magnetic unit cells, and $N$ is the number of
sites in the atomic lattice.  We note that, assuming an $L_x\times L_y$
magnetic unit cell, the number of atomic sites is just $L_x L_y N_M$, so we
may simply write $\chi_{\text{AF}}(\bq,\delta\br-\delta\br',\omega)=$
\begin{equation}
\frac{1}{4}\frac{1}{L_x L_y}\sum_{\bG}\sum_i\frac{\alpha\xi^2\mu_B e^{i(\bq+\bG)\cdot(\delta\br-\delta\br')}}{1+(\bq+\bG-\bQ_i)^2+i\omega/\omega_{\text{SC}}}.\label{AFsusc_comp}
\end{equation}
So far, we have not introduced a new model; we only rewrote the original in a
more complicated form.  We will now modify this model to introduce vortex effects.
For simplicity, we will assume only one vortex per unit cell.  We model these
effects by simply introducing step functions into the susceptibility to restrict
the antiferromagnetic correlations to within a distance equal to the superconducting
coherence length $\xi_{\text{SC}}$ from the center of the vortex.  The result is
$\chi_{\text{AF}}(\bq,\delta\br,\delta\br',\omega)=$
\begin{eqnarray}
\frac{1}{4}\frac{1}{L_x L_y}\sum_{\bG}\sum_i\frac{\alpha\xi^2\mu_B e^{i(\bq+\bG)\cdot(\delta\br-\delta\br')}}{1+(\bq+\bG-\bQ_i)^2+i\omega/\omega_{\text{SC}}}\times\cr
\times\theta(\xi_{\text{SC}}-\delta r)\theta(\xi_{\text{SC}}-\delta r').
\end{eqnarray}
We now turn our attention to finding the spin-lattice relaxation rate from this
susceptibility.  It can be shown that, if we assume that this relaxation is due
to the hyperfine interaction \eqref{HFIntV}, then the spin-lattice relaxation rate
will be
\begin{eqnarray}
\frac{1}{T_1(\br)}=\gamma_e^2\gamma_n^2\hbar^3 k_B T\sum_{\br'\br''}A(\br-\br')A(\br-\br'')\times\cr
\times\lim_{\omega\rightarrow 0}\frac{\Di\chi_{+-}(\br',\br'',\omega)}{\hbar\omega},
\end{eqnarray}
where $\Di{f(\ldots,\omega)}$ is a ``discontinuity'' operator, defined as
\begin{equation}
\Di{f(\ldots,\omega)} = \frac{f(\ldots,\omega+i0^+)-f(\ldots,\omega-i0^-)}{2i}.
\end{equation}
We may rewrite the above formula for a susceptibility of the form we are working with.
By introducing the appropriate Fourier transforms, we eventually arrive at the desired
result,
\begin{eqnarray}
\frac{1}{T_1(\br)}=\gamma_e^2\gamma_n^2\hbar^3 k_B T\frac{1}{L_x^2 L_y^2 N_M}\times\cr
\times\sum_{\delta\br'\delta\br''}\sum_{\bG_1\bG_2}\sum_{\bk}A(\bk+\bG_1)A^{\ast}(\bk+\bG_2)e^{i(\bG_1-\bG_2)\cdot\br}\times\cr
\times e^{-i(\bk+\bG_1)\cdot\delta\br'}e^{i(\bk+\bG_2)\cdot\delta\br''}\lim_{\omega\rightarrow 0}\frac{\Di\chi_{+-}(\bk,\delta\br',\delta\br'',\omega)}{\hbar\omega},
\end{eqnarray}
where $\delta\br'$ and $\delta\br''$ are summed over the entire unit cell, $\bk$ is summed
over the entire magnetic Brillouin zone, and $\bG_1$ and $\bG_2$ are summed over the same
set of vectors as $\bG$ in Equation \eqref{AFsusc_comp}.

The rest of our work was done numerically.  We used the experimentally-determined parameters
given by Barzykin and Pines for the susceptibility for YBa$_2$Cu$_3$O$_{6.63}$ \cite{Barzykin1995}.
We first performed a numerical calculation of the rates for $\br=0$ for both copper and oxygen
without vortices.  The form factors we used are
\begin{eqnarray}
A_{\text{Cu}}(\bk)=A+2B(\cos{k_x a}+\cos{k_y a}) \\
A_{\text{O},x}(\bk)=2\cos{\tfrac{1}{2}k_x a}(C_1+2C_2\cos{k_y a})\label{FF_Ox} \\
A_{\text{O},y}(\bk)=2\cos{\tfrac{1}{2}k_y a}(C_1+2C_2\cos{k_x a})\label{FF_Oy},
\end{eqnarray}
where $a$ is the atomic lattice spacing and the parameters, $A$, $B$, $C_1$,
and $C_2$ are those given by Zha, Barzykin, and Pines \cite{Zha1996}.  Because
of the fact that the momentum-space points $(k_x,k_y)$ and $(k_y,k_x)$ are both
present in the sums in our formulas, Equations \eqref{FF_Ox} and \eqref{FF_Oy}
should give the same result.  We did this using both Equations \eqref{AFsusc_simp}
and \eqref{AFsusc_comp} to check our formulas.  The temperature range we examined
was $70\text{ K}\leq T\leq 300\text{ K}$.  We found that the contributions to the
rates for both Cu and O decreased with increasing temperature, and that the rates
for O were several orders of magnitude smaller than for Cu; such suppression of the
rate for O compared to Cu has been reported before by Mila and Rice\cite{MilaRice1989}.
We then repeated this calculation, this time including vortices.  We set the
superconducting coherence length $\xi_{\text{SC}}=2a$.  We found that the temperature
dependence of both rates was qualitatively the same as before, but that the rates
were actually enhanced; the copper rates by an entire order of magnitude and the
oxygen rates by a factor of about 4.  We believe that this is due to the fact that,
by imposing the distance cutoff, we removed contributions to the relaxation rate
that would have reduced the rate.  Based on this simplified model, we therefore
expect that, not only is the filtering effect still present in the mixed state,
but it is, in fact, enhanced.

It goes without saying then that this filtering effect would
therefore make oxygen NMR more sensitive to the pair creation and
annihilation (PCA) processes than copper NMR. This is because the
effect of antiferromagnetic correlations is much less for oxygen
than for copper, meaning that the PCA processes will dominate in
oxygen.

\section{Conclusions}
In this work, we argued that it is possible to explain the
broadening of the line shape and the upturn in the spin-lattice
relaxation rate with decreasing temperature observed experimentally
\cite{MitrovicNature,MitrovicPRB2003} without introducing
antiferromagnetic correlations.  The line shape broadening can be
explained, at least in part, by noting that the Knight shift varies
with position in the lattice in the vortex state.  This position
dependence leads to each nucleus having a different resonance
frequency, and therefore to a broadened line shape.  The upturn in
the relaxation rate can be explained as due to a second relaxation
process, namely creation and annihilation of pairs of spin-up
quasiparticles, that appears when a magnetic field is applied, and
this process dominates at low temperatures.  We do not wish to claim
that AF correlations do not exist in YBCO, only that certain features
of the NMR data once attributed to such correlations can be explained
without them; in fact, there is other evidence for the existence
of such correlations, namely neutron scattering data\cite{Vaknin2000}.
As we argued in Section IV, even in the presence of AF correlations, the spin-lattice
relaxation rates for O will not be greatly affected by them due to the
form factor.

Based on the above arguments we expect that, once the vortex lattice
melts and the system enters a vortex liquid phase, the NMR lines
sharpen due to motional narrowing\cite{reyes1997,kittelBook}. At the
same time, we expect that the spin lattice relaxation rate, $1/T_1$,
is determined by the faster rates and that the low $T$ upturn
persists in the vortex liquid.

This picture, and the density of states shown in Fig.\ref{fig:DOS},
also predict that if an experiment is performed in a clean thin film
with a well-ordered vortex lattice in which the perpendicular
component of the $B$-field is kept fixed, while changing the
magnitude of the total $\bB$, {\it quantum-like oscillations} in
$1/T_1$, due to the oscillations of the density of states {\it in
energy}, would be observed.

While our calculated line shapes have about the same width as the
experimental shape for the $42\text{ T}$ case, the shapes for the
$13\text{ T}$ case have different widths.  One possible contributing
factor to this discrepancy is the fact that we neglected the variation
of the magnetic field and the pairing amplitude over a unit cell.  We
expect the magnetic field to vary more strongly in the $13\text{ T}$
case than in the $42\text{ T}$ case because the vortices are further
apart in the $13\text{ T}$ case.  In fact, in the $42\text{ T}$ case,
the distance between the two vortices in a unit cell is about $10\%$
of the penetration depth, while, in the $13\text{ T}$ case, this distance
is about $23\%$ of the penetration depth.  This variation will introduce
further broadening, which will be greater at $13\text{ T}$ than
at $42\text{ T}$, consistent with our findings.

We also notice that the ``tails'' on our calculated curves are different
in length than those of the experimental curves.  We believe that this,
once again, is due to the fact that we neglected the variation of the
pairing potential over a unit cell.  In reality, the order parameter should
be lower in magnitude near the vortex cores because these regions are where
superconductivity is beginning to break down.  This means that we expect
our calculated line shapes to be more accurate in the lower internal field
regions than in the high internal field regions.

Finally, we note that the peak in our curve at $42\text{ T}$ 
is split in two, as opposed to the single peak seen in the experimental
data\cite{MitrovicNature,MitrovicPRB2003}.  This suggests that there is another
broadening mechanism at work besides that due to the finite width of the normal-state
line shape because such broadening can wash out the ``split'' peak so that only a
single peak appears.  One such possibility is the presence of impurities.

We are able to obtain good fits of our calculated temperature dependence of the
spin-lattice relaxation rates to the experimental data using the internal magnetic
field as our {\it only} fitting parameter.  We note, however, that the values
of the internal magnetic fields giving us our best-fit curves on the line shape
do not quite match the experimental results.  In the experiment, the region away
from the core was in the vicinity of the peak in the line shape\cite{MitrovicPRB2003}.
However, the positions of the corresponding theoretical curves do not quite fall
on the peak; rather, they are away from it.  It is possible that this discrepancy
may be due, in part, to our approximations in solving the Bogoliubov-de Gennes
equation, and that more realistic modeling of the vortex core is necessary to
account for this.

\acknowledgements
We wish to thank Profs. Mitrovi\'c, Kivelson, Chakravarty and Te\v
sanovi\'c for discussion. OV would also like to thank the Aspen
Center for Physics, where part of this work was completed, for
hospitality.

\appendix
\section{Details of Derivation of Spin-Lattice Relaxation Rate}
Here, we go into some more detail about how we found the different terms
in the spin-lattice relaxation rate.  As we mentioned earlier, when we
rewrite the spin raising and lowering operators in terms of the Bogoliubov
quasiparticles, we obtain 16 terms, though only 6 give non-zero contributions.
One term we find that contributes to the SF process, suppressing factors
of $u$ and $v$ that occur, is
\begin{eqnarray}
\sum_{QQ'}\sum_{\bk n\bk' n'}\sum_{\bq m\bq' m'}\left <Q'\right |\hat{\gamma}_{\bk n\downarrow}^{\dag}\hat{\gamma}_{\bk' n'\uparrow}\left |Q\right >\left <Q\right |\hat{\gamma}_{\bq m\uparrow}^{\dag}\hat{\gamma}_{\bq' m'\downarrow}\left |Q'\right > \cr
\times\delta(E_{Q'} - E_{Q}).
\end{eqnarray}
We note that the only terms that will give non-zero contributions are those
for which $\bk = \bq'$, $\bk' = \bq$, $n = m'$, and $n' = m$.  We may then
write
\begin{eqnarray}
\sum_{QQ'}\sum_{\bk n\bk' n'}\left <Q'\right |\hat{\gamma}_{\bk n\downarrow}^{\dag}\hat{\gamma}_{\bk' n'\uparrow}\left |Q\right >\left <Q\right |\hat{\gamma}_{\bk' n'\uparrow}^{\dag}\hat{\gamma}_{\bk n\downarrow}\left |Q'\right > \cr
\times\delta(E_{Q'} - E_{Q}).
\end{eqnarray}
We now note that the only non-zero matrix elements will be those in which
the state $Q'$ is obtained from the state $Q$ by scattering a particle from
the state with crystal wave vector $\bk'$, band index $n'$, and spin up
into the state with wave vector $\bk$, band index $n$, and spin down.  This
means that the energy difference between the two states is just
$E_{Q'}-E_{Q}=E_{\bk n\downarrow}-E_{\bk' n'\uparrow}=E_{\bk n}-E_{\bk' n'}+2h$.
This energy difference is independent of the exact many-particle states $Q$
and $Q'$, so we may rewrite the sum on these states as a trace:
\begin{eqnarray}
\sum_{Q'}\sum_{\bk n\bk' n'}\left <Q'\right |\hat{\gamma}_{\bk n\downarrow}^{\dag}\hat{\gamma}_{\bk' n'\uparrow}\hat{\gamma}_{\bk' n'\uparrow}^{\dag}\hat{\gamma}_{\bk n\downarrow}\left |Q'\right > \cr
\times\delta(E_{\bk n} - E_{\bk' n'} + 2h)
\end{eqnarray}
We may now employ the anticommutation relations among the quasiparticle
operators to rewrite the above in terms of number operators:
\begin{eqnarray}
\sum_{Q'}\sum_{\bk n\bk' n'}\left <Q'\right |\hat{n}_{\bk n\downarrow}(1 - \hat{n}_{\bk' n'\uparrow})\left |Q'\right > \cr
\times\delta(E_{\bk n} - E_{\bk' n'} + 2h)
\end{eqnarray}
Upon taking the thermal average, the trace becomes a product of
Fermi functions:
\begin{eqnarray}
\sum_{\bk n\bk' n'}f(E_{\bk n\downarrow})(1 - f(E_{\bk' n'\uparrow})) \cr
\times\delta(E_{\bk n} - E_{\bk' n'} + 2h)
\end{eqnarray}
We will also show the calculation for the term resulting from the
pair creation and annihilation term because it will differ slightly
from the calculation given above.  This term (again suppressing
actors of $u$ and $v$) is
\begin{eqnarray}
\sum_{QQ'}\sum_{\bk n\bk' n'}\sum_{\bq m\bq' m'}\left <Q'\right |\hat{\gamma}_{\bk n\uparrow}\hat{\gamma}_{\bk' n'\uparrow}\left |Q\right >\left <Q\right |\hat{\gamma}_{\bq m\uparrow}^{\dag}\hat{\gamma}_{\bq' m'\uparrow}^{\dag}\left |Q'\right > \cr
\times\delta(E_{Q'} - E_{Q}).
\end{eqnarray}
In this case, there are two ways to ``match'' the operators; we may
either let $(\bk, n) = (\bq, m)$ and $(\bk', n') = (\bq', m')$ or
let $(\bk, n) = (\bq', m')$ and $(\bk', n') = (\bq, m)$.  We thus
obtain
\begin{eqnarray}
\sum_{QQ'}\sum_{\bk n\bk' n'}(\left <Q'\right |\hat{\gamma}_{\bk n\uparrow}\hat{\gamma}_{\bk' n'\uparrow}\left |Q\right >\left <Q\right |\hat{\gamma}_{\bk n\uparrow}^{\dag}\hat{\gamma}_{\bk' n'\uparrow}^{\dag}\left |Q'\right > + \cr
\left <Q'\right |\hat{\gamma}_{\bk n\uparrow}\hat{\gamma}_{\bk' n'\uparrow}\left |Q\right >\left <Q\right |\hat{\gamma}_{\bk' n'\uparrow}^{\dag}\hat{\gamma}_{\bk n\uparrow}^{\dag}\left |Q'\right >) \cr
\times\delta(E_{Q'} - E_{Q}).
\end{eqnarray}
The only non-zero matrix elements in this case are those in which the
state $Q'$ is obtained from the state $Q$ by destroying two quasiparticles
with spin up, one with wave vector $\bk$ and band index $n$ and one
ith wave vector $\bk'$ and band index $n'$.  This means that the energy
difference $E_{Q'}-E_{Q}=-E_{\bk n\uparrow}-E_{\bk' n'\uparrow}=-E_{\bk n}-E_{\bk' n'}+2h$.
The rest of the derivation proceeds as before, and we eventually obtain
\begin{eqnarray}
\sum_{\bk n\bk' n'}(1 - f(E_{\bk n\uparrow}))(1 - f(E_{\bk' n'\uparrow})) \cr
\times\delta(E_{\bk n} + E_{\bk' n'} - 2h).
\end{eqnarray}


\end{document}